\documentclass[letterpaper]{article} 
\usepackage{aaai23}  
\usepackage{times}  
\usepackage{helvet}  
\usepackage{courier}  
\usepackage[hyphens]{url}  
\usepackage{graphicx} 
\usepackage{natbib}  
\usepackage{caption} 
\usepackage{url}
\usepackage{latexsym}
\usepackage{array}
\usepackage{todonotes}
\usepackage{subfigure} 
\usepackage{algorithm}
\usepackage{algorithmic}
\usepackage{makecell}
\usepackage{booktabs}
\usepackage{listings}
\usepackage{arydshln}
\usepackage{inconsolata}
\usepackage{adjustbox}
\usepackage{amssymb,amsmath,array}
\usepackage{multirow}
\usepackage{comment}
\usepackage{balance}
\usepackage{algorithm}
\usepackage{algorithmic}
\newcommand{\xiw}[1]{\textcolor{black}{#1}}
\newcommand\blfootnote[1]{%
  \begingroup
  \renewcommand\thefootnote{}\footnote{#1}%
  \addtocounter{footnote}{-1}%
  \endgroup
}
\usepackage{newfloat}
\usepackage{listings}
\DeclareCaptionStyle{ruled}{labelfont=normalfont,labelsep=colon,strut=off} 
\floatstyle{ruled}
\newfloat{listing}{tb}{lst}{}
\floatname{listing}{Listing}

\title{Task2KB: A Public Task-Oriented Knowledge Base}
\author{
Procheta Sen$^\ast$\textsuperscript{1}, Xi Wang$^\ast$\textsuperscript{2}, Ruiqing Xu\textsuperscript{3}, Emine Yilmaz\textsuperscript{2} }
\affiliations{
 \textsuperscript{1}University of Liverpool, Liverpool, UK \\
 \textsuperscript{2}University College London, London, UK \\
 \textsuperscript{3}University of Manchester, Manchester, UK  \\
  procheta.sen@liverpool.ac.uk, xi-wang@ucl.ac.uk,\\
ruiqing.xu@outlook.com, emine.yilmaz@ucl.ac.uk\\}
 
\usepackage{bibentry}

\begin{document}

\maketitle
\blfootnote{\hspace{-0.7em}$^\ast$These authors contributed equally to this work.}

\begin{abstract}
Search engines and conversational assistants are commonly used to help users complete their every day tasks such as booking travel, cooking, etc. 
While there are some existing datasets that can be used for this purpose, their coverage is limited to very few domains. In this paper, we propose a novel knowledge base, ‘Task2KB’, which is constructed using data crawled from WikiHow, an online knowledge resource offering instructional articles on a wide range of tasks. Task2KB encapsulates various types of task-related information and attributes, such as requirements, detailed step description, and available methods to complete tasks. Due to its higher coverage compared to existing related knowledge graphs, Task2KB can be highly useful in the development of general purpose task completion assistants. 

\end{abstract}
\section{Introduction}
Task-oriented conversational systems~\cite{Kim2020,task2} have recently become popular to support users in completing many tasks. A well-performed conversational agent requires a comprehensive knowledge about the every stage of a given task to make proper responses or ask useful clarification questions. 
\xiw{Therefore to develop effective conversational models, existing task-based conversation models have used various datasets, like MultiWOZ~\cite{zang2020multiwoz}, MSDialog~\cite{InforSeek_User_Intent} and UDC~\cite{UDC}. However, a common limitation of these datasets is their low coverage of topics from multiple domains. For example,} MultiWOZ is limited to the use of data from few domains (e.g. restaurant booking, hotel finding, taxi booking). 
As a consequence, it is challenging to learn and train a conversational agent in supporting tasks other than the ones mentioned in existing datasets.

\looseness -1 \xiw{In existing literature,}  there also exists some works which attempted to develop a task-based knowledge base~\cite{chu,pareti2014integrating}. For example,~\citet{chu} created a knowledge base named \textit{How2KB} from WikiHow descriptions. WikiHow\footnote{\url{https://wikihow.com}} is an online knowledge resource, which offers instructional descriptions of a large number of tasks from different domains (e.g.\ Food and Entertainment, Travel, Holiday, Health). For each task, How2KB provides its location, time, participating agent and participating object, previous task, next task and sub-tasks. However, How2KB does not provide a full access to the task-related information and missing a convenient way for retrieving task information from particular domains. The information stored in How2KB is also noisy. 
To alleviate the above mentioned limitations, we develop a task-oriented knowledge base from WiKiHow in an automated way. 
Along with such a knowledge base, we further create a system named `Task2KB'\footnote{~\url{https://www.task2kb.uk}}. \xiw{Task2KB enables the access to a comprehensive description of many tasks that belongs to the listed domains.} 
In addition, our Task2KB system also allows users to interact with our interface to retrieve the related tasks to a given task. 

\section{Knowledge Base Construction} \label{sec:kb}
\begin{figure}
    \centering
    \includegraphics[width=0.23\textwidth]{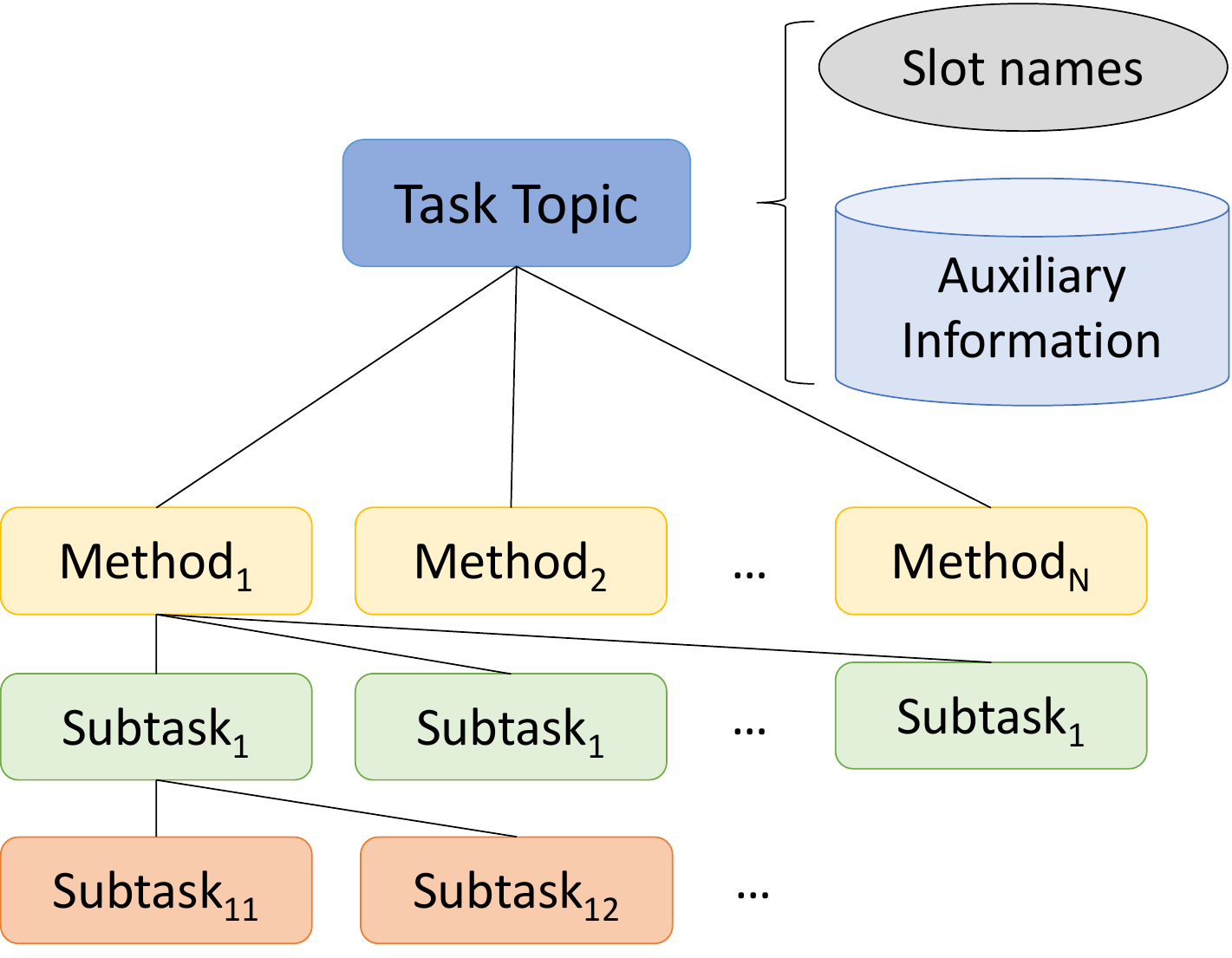}    
    \caption{Structure of the knowledge base.}
    \label{fig:kbstruct}
\end{figure}

\looseness -1 In this section, we first describe the structure of our knowledge base. Next, we provide a detailed description of calculating the estimated quality of task articles as well as the identification of slot names of tasks. For each task stored in our knowledge base, we have its mandatory and optional attributes. Figure~\ref{fig:kbstruct} shows the structure of our knowledge base. The mandatory attributes for each task are the available methods \xiw{and the corresponding} sub-task for completing a task. A sub-task is similar to a task, \xiw{which has an associated} goal to be accomplished for a successful completion of its parent task. In our research scope, we consider each step information of a task as its sub-task. In WikiHow task articles, step descriptions \xiw{are in distinct lengths} and may \xiw{further involve} multiple sub-steps. In this scenario, a sub-task can also have its own sub-task \xiw{(i.e.\ sub-sub-task)} as shown in Figure~\ref{fig:kbstruct}. 

In our knowledge base, we also provide some optional attributes for tasks (i.e.\ the auxiliary information in Figure~\ref{fig:kbstruct}). Example auxiliary attributes are tips, reviews, summary, questions, slot names, quality score. Every auxiliary attribute of a task can be directly crawled from the corresponding WikiHow page, the except slot information and quality score. 
Next, we describe about quality estimation and slot information extraction in detail. 

\xiw{\textbf{Article Quality Estimation of Tasks}}\label{sec:quality}
As denoted in~\cite{wang2020negative}, human-labelled information is frequently biased due to its insufficient or limited time of posting. 
Therefore, to fairly present the article quality and usefulness of a task article, we leverage seven types of  features including the articles' recency, their normalised number of views, normalised number of votes, their introduction, summary, description as well as their detailed step information to estimate the quality of an article. We apply BERT~\cite{BERT}, to encode the text with the introduction, summary and description, while leveraging the LongFormer model~\cite{longformer} to transform the text of the step information into embedding due to its frequent long length. Then, we use the averaged score of each article as ground truth to train a deep neural model with multiple convolutional layers
for the article quality prediction. We represent the quality of task articles as a score which ranges in [0, 1]. A higher quality article has its quality score closer to 1. 

\textbf{Finding Slot Names of a Task}\label{sec:attr}
Slots in dialogues have been commonly used by various conversational systems to address dialogue state tracking, which is a key component of a dialogue system~\cite{ye2021slot}. For example, some possible slots for a restaurant booking task are `restaurant location', `rating' and `price'. 
Existing task oriented conversational datasets, like MultiWOZ, have manually curated slots for each task. However, it is challenging to manually create slots for a large number of tasks. To address the above mentioned issue, in Task2KB, we propose an automated method to prepare slots for a task. For a given task, we first extract the entities from the task description. Then, for each entity, we collect its corresponding attributes from the Wikipedia\footnote{\url{https://wikipedia.org}} page. As a result of this, we obtain a set of potential attributes for each task. Now, for each task we represent it with the embeddings of the corresponding entities. 
We compute the embeddings of both entities and attributes with a pre-trained BERT model\footnote{\url{https://huggingface.co/docs/transformers/model_doc/bert}}. 
Next, we rank the attributes of a task based on the cosine similarity computed between the corresponding task and its attributes. For each task, we use the top $K$ similar attributes as the possible slot names of a given task. Example slots for the task `How-To-Ship-Food' are `packaging', 'fridging method', `shipping policy'.

\begin{figure}
    \centering
    \includegraphics[width=0.35\textwidth]{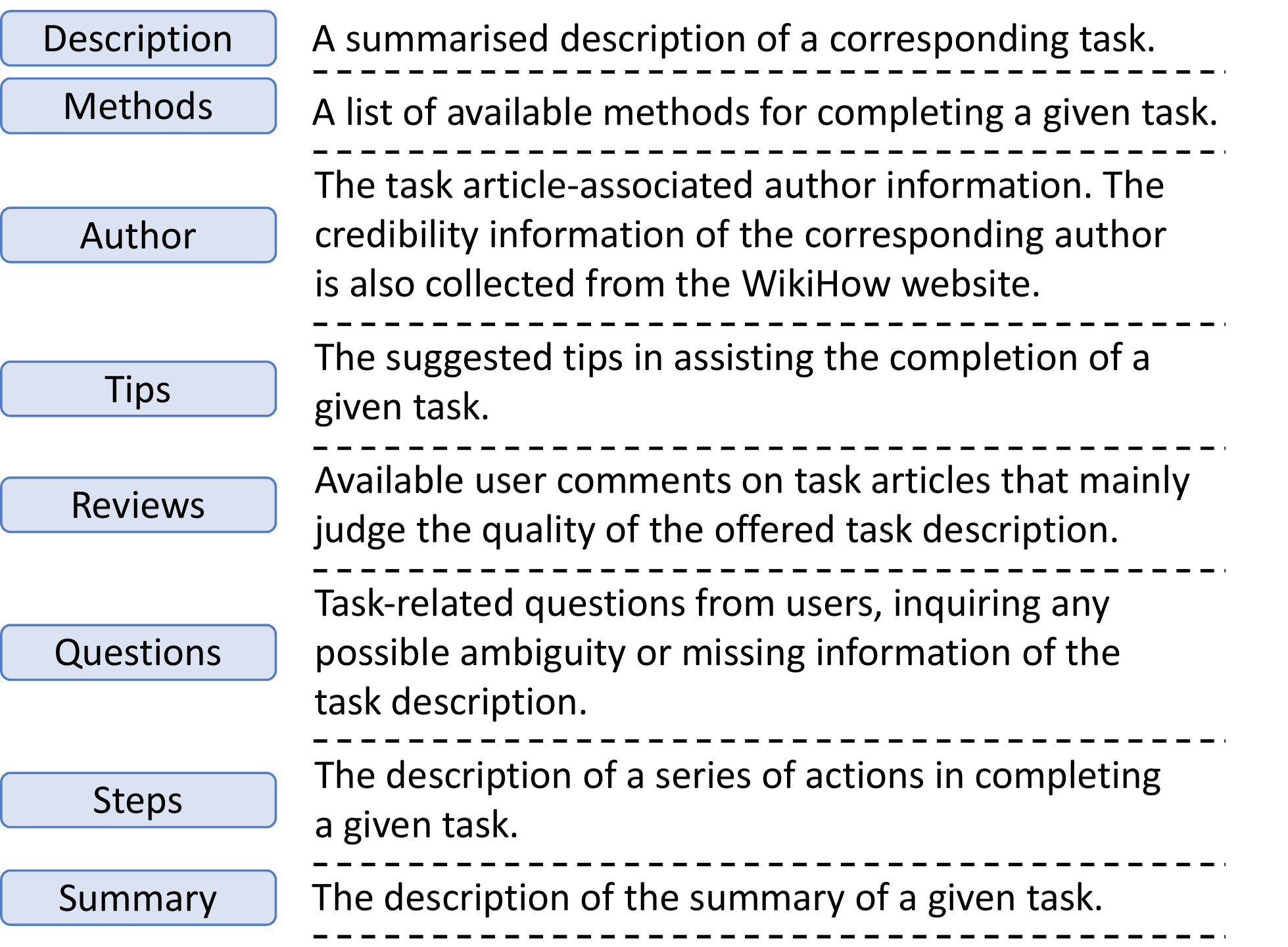}
    \caption{Available information summary in Task2KB.}
    \label{fig:information_summary}
\end{figure}

\textbf{Dataset Collection}\label{sec:dataset}
We crawled data from the WikiHow website. WikiHow, has in total $13$ different categories / domains. 
So far, in Task2KB, we first crawled data for `Food and Entertainment', `Travel', `Holiday', and `Health', which results in total $44,831$ documents in our knowledge base. In the near future, the Task2KB will be extended by including task articles from additional domains. 
In Figure~\ref{fig:information_summary}, we also describe a list of available information from WikiHow for every task in our Task2KB system.

\section{Functionalities of Task2KB API} \label{sec:features}

\textbf{Finding Potential Tasks from a Task Description}\label{sec:feature1}
In Task2KB, given a textual content, we provide a ranked list of tasks which can be associated with it. 
We use a Field based retrieval model~\cite{Kim} to find potential tasks corresponding to a text phrase. Equation~\ref{eq:field} shows the expression for score computation formula between a document $D$ and a query $Q$ in Field based retrieval model. In Equation~\ref{eq:field}, $n$ is the number of fields in an index and $m$ is the number of query terms and $F_j$ is a particular field. In our research scope, the number of fields used to search is $3$ (i.e. step information, task description, task summary). 
\begin{equation}
     S(Q,D)=\prod_{i=1}^m\sum_{i=1}^nP(F_j|q_i, R)P(q_i|F_j, D)
     \label{eq:field}
\end{equation}

\textbf{Finding Detailed Information About a Task}
Task2KB enables users' search requirements on a specified domain. After that, in Task2KB, we also allow users to retrieve the overall information of their target tasks. Such information includes the task title, task domain and the set of sub-tasks related to the task. 
We also aggregate many advanced information for a given task, which can also be accessed from our Task2KB. Such advanced information includes the author of the task article, slot names corresponding to the task, questions asked about the task, tips related to the task, reviews corresponding to the task description. 

\section{Conclusion and Future Work}

In this paper, we introduce a task based knowledge base, \textit{Task2KB}. Task2KB provides rich information about tasks from various domains, which could enhance many task-oriented tasks, such as conversational systems and task-oriented recommendations. In the future, we aim to leverage the available data resources in Task2KB to develop effective task-oriented conversational models.

\bibliography{main}

\end{document}